# The influence of chromosomal inversions on genetic variation and clinal patterns in genomic data of *Drosophila melanogaster*

Martin Kapun[1]

[1] Natural History Museum of Vienna, Vienna, Austria

**Abstract**

Chromosomal inversions are structural mutations resulting in the reversal of the gene order along the corresponding genomic region. Due to their influence on recombination patterns, they can have a major influence on genetic variation and the evolutionary process. Accordingly, inversions can act as supergenes that keep together co-adapted gene complexes that form the genetic basis of many complex phenotypes in diverse organisms. In this book chapter, I will present an analysis pipeline to investigate the influence of two common cosmopolitan inversion, *In(2L)t* and *In(3R)Payne*, on genome-wide genetic variation and differentiation in world-wide populations of the vinegar fly *Drosophila melanogaster.* We will use single-individual and pooled resequencing data in combination with population genomics analysis tools to explore the impact of these two inversions on genetic variation, population structure, and clinal variation in natural populations.

## 1       Introduction

Chromosomal inversions are structural mutations that result in the reorientation of the gene order in the affected genomic region *[1–3]*. The reversal of synteny impedes homologous pairing in heterokaryotypic chromosomes, i.e., a chromosome pair, where one copy is inverted (INV) and the other of standard (ST) arrangement, and leads to loop structures inside the chromosomal region spanned by the inversion *[4]*. As shown in Figure 1, these characteristic inversion loops can even be examined under light microscopes in giant polytene chromosomes, which represent thousand-fold replicated chromatids within the nucleus that can be found, for example, in the salivary glands of many drosophilid larvae *[5–7]*. These structures allowed investigating the influence of inversions on recombination patterns in the early days of genetics research, approximately 100 years ago, in the vinegar fly *D. melanogaster [e.g., 8, 9]* and subsequently also in *D. pseuoobscura [7, 10, 11]*, making these structural polymorphisms one of the first mutations ever to be directly studied. Inversions can either be the result of erroneously mended double strand breaks *[12]* or of ectopic recombination among repetitive and palindromic sequences, for instances tRNAs, ribosomal genes *[13]* and transposable elements *[TEs; 14]*. Accordingly, the breakpoints of inversion polymorphisms, which can range from less than a thousand to several million base pairs in length, are often enriched for these repetitive sequences  and predominantly occur in "weak" spots of the genome that are prone to breakage *[15]* and that can even be shared across distantly related taxa *[16]*. The prevalence of inversions that either include or exclude the centromere (pericentric vs. paracentric inversions) can vary dramatically, even among closely related taxa, which may be linked to varying numbers of TEs in different genomes. For example, in contrast to the vinegar fly *D. melanogaster*, which harbors inversions common in many worldwide populations, its sister taxa *D. mauritiana* and *D. simulans* are basically inversion-free *[17–19]*.

The primary evolutionary effect of inversions is a strong suppression of recombination with standard arrangement chromosomes, since crossing-over in heterokaryotypes within the inverted region results in unbalanced gametes that are non-viable *[4, 10]*. While recombination with paracentric inversions results in acentric and dicentric gametes *[20]*, crossing-over with pericentric inversions can cause large-scale duplications and deletions in the recombination products. As a result, both the ancestral standard (ST) and the derived inverted (INV) karyotype evolve largely independently *[21–23]*. However, the suppression of recombination is not perfect and two processes may lead to rare genetic exchange - so called "gene flux" - across karyotypes *[23]*. In the first process two recombination events (double recombination) happen at the same time within the inverted region and may result in viable recombinant gametes. However, the synaptonemal complexes, which initiate crossing over, can only form when the homologous chromatids are fully paired *[24]*. Thus, rare double recombination events never occur in the proximity of the inversion breakpoints and consequently lead to a gradual increase of gene flux probability towards the inversion center *[25]*. Conversely, in the second process gene conversion, which results from repair of DNA double-strand breaks and does not depend on paired chromatids, can lead to rare genetic exchange across the whole inverted region *[26]*.

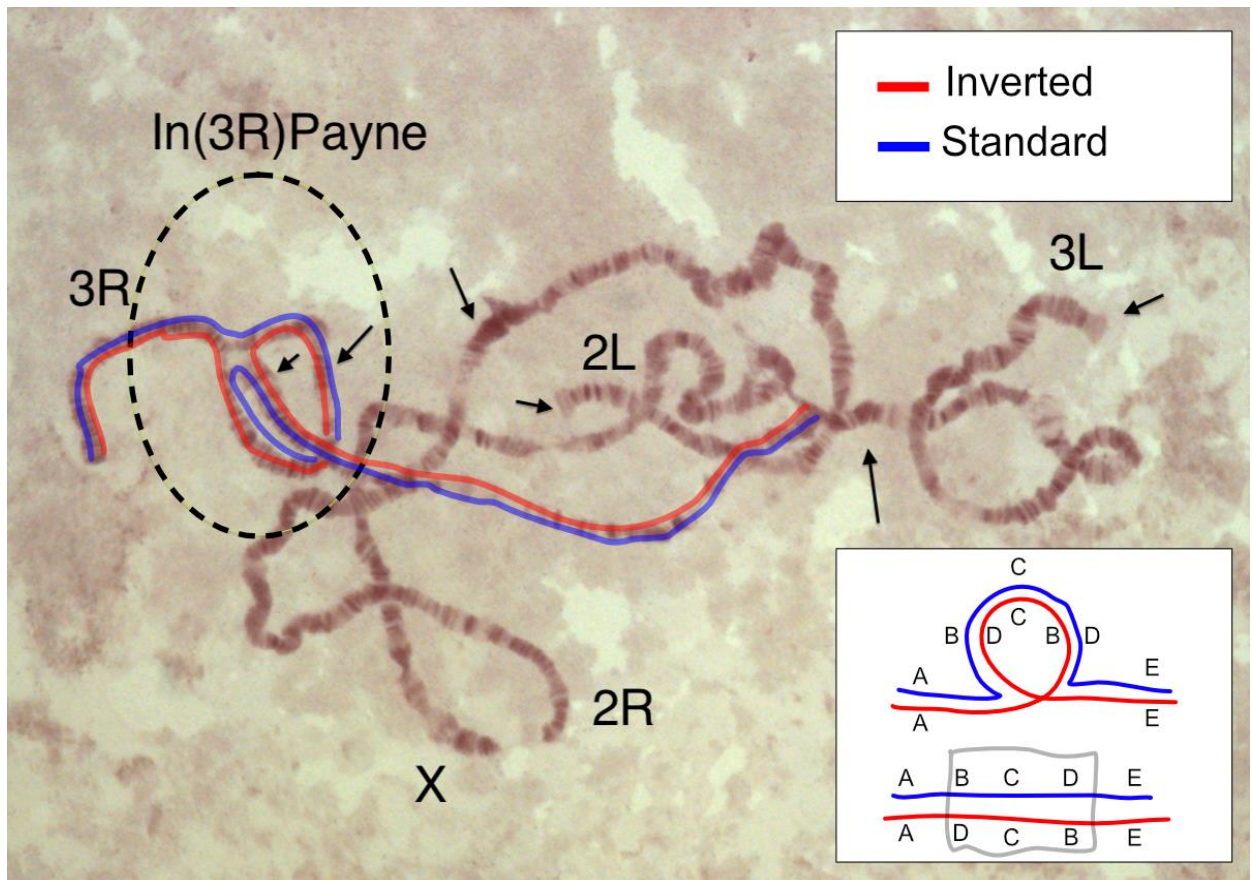

**Figure 1**: Cytological preparation of polytene chromosomes from the salivary glands of a third instar *Drosophila melanogaster* larva which is heterozygote for the *In(3R)Payne* inversion on the right arm of the third chromosome (*3R*). The arrows pinpoint banding patterns that are diagnostic for the chromosomal arms, the dashed-lined ellipse indicates the inversion and the red and blue lines indicate the orientation of the two paired sister chromatids (inverted and standard, respectively) with different karyotype that form a characteristic inversion loop. The insert in the bottom right

corner shows a schematic representation of the inversion loop, where the letters represents genes along the chromosome which should illustrate that the region spanned by "BCD" is inverted in the red haplotype as highlighted by a grey box on top of the linear alignment of the two karyotypes at the bottom of the insert.

When a new inversion arises, it captures a single haplotype of the ancestral standard arrangement. The evolutionary fate of the new inversion is initially determined by genetic drift and by the fitness of the captured haplotype relative to the rest of the population *[1, 22, 27]*. Subsequently, divergence among the karyotypes builds up continuously due to the accumulation of novel mutations. However, gene flux away from the breakpoints keeps homogenizing the genetic variation of the remaining genome *[28]* and can lead to a pattern of sequence divergence that resembles a "suspension bridge" in inversions that are sufficiently old to harbor many novel mutations *[29]*. Accordingly, inversions may strongly influence the patterns of genetic variation in the corresponding genomic region *[22, 27, 30]*.

Particularly large chromosomal inversions are considered to play an important evolutionary role given their impact on genetic processes. These effects range from (1) positional shifts of genes within a chromosome, which may perturb their expression patterns *[e.g., 31, 32]*; (2) inviable recombination products in heterozygotes, as explained above; (3) putative pseudogenization in the breakpoint regions *[e.g., 33]* when the inversion disrupts a previously functional gene. While these effects can be deleterious and lead to a rapid loss of the inversion, they may also provide strong adaptive effects. In support of an adaptive value of inversions, we find overwhelming evidence that many inversions are common and contribute to adaptive phenotypic variation *[for comprehensive reviews, see 1–3, 29, 30, 34–37]*. Given their ability to suppress recombination, inversion can keep co-evolved genes together and protect these against maladaptive recombination and thus act as so-called supergenes *[38, 39]*. Accordingly, inversions commonly form the genetic basis of complex phenotypic variation. Prominent examples of phenotypes influence by inversions include male morphs in the ruff, a beautiful wader bird which is characterized by discrete behavioral and morphological differences in males *[40–42]*, mimicry in wing coloration in *Heliconius* butterflies *[43, 44]*, ecotypes in monkeyflower of the genus *Mimulus* adapted to different environmental conditions *[45, 46]*, and shell morphs of *Littorina saxatilis*, which are snails adapted to different locations and conditions along the marine intertidal shoreline *[47]*.

Different selection mechanisms may directly or indirectly favour inversions and lead to an increase in their frequency *[3, 29, 35, 36, 48]*. Inversions can, for example, accumulate beneficial mutations around their breakpoints and increase in frequency due to strong linkage disequilibrium (LD) with the positively selected alleles *[1]*. Alternatively, inversions may pick up a locally adapted haplotype through gene flux and then become highly beneficial because they protect the local haplotype against maladaptive recombination with deleterious haplotypes that are introduced to the population through migration *[37, 49, 50]*. Inversions may also spread because they keep together epistatic combinations of beneficial alleles inside the inversion even in absence of migration load *[51–54]*. In this case, inversions confer a selective advantage by suppressing recombination between beneficial allelic combinations. The beneficial effect of the inversion is thus highest when it segregates at intermediate frequencies and the proportion of heterozygotes is thus maximized in a population. These are thus often assumed to be under balancing selection, which maintains them at intermediate frequencies *[reviewed in 1, 30]*. Consistent with this hypothesis, inversions are commonly found to occur at intermediate frequencies in many populations, such as the common cosmopolitan inversions *In(2L)t* and *In(3R)Payne* in *D. melanogaster [19, 35]*. Moreover, the observation of clinal patterns along different environmental gradients on multiple continents further indicate that inversions play a key evolutionary role and are maintained due to ecological selection *[e.g., 55–59]*.

In this book chapter, I will present a bioinformatics analysis pipeline which allows to assess the influence of inversions on genetic variation in natural populations. As an example we will focus on the vinegar fly *D. melanogaster*, which is characterized by seven chromosomal inversions that are commonly found in many populations world-wide *[19]*. Using genomic data from different sources and a broad range of bioinformatics analyses tools, we will study the two aforementioned inversions *In(2L)t* and *In(3R)Payne*, which originated in Africa *[31, 60]* and are now common cosmopolitan inversions, in a population sample from Zambia, and investigate their effect on genetic variation and differentiation. We will then identify the single nucleotide polymorphisms (SNPs) in the proximity of the inversion breakpoints that are fixed for different alleles in the inverted and standard chromosomal arrangements *[61]* using samples with known karyotypes. Using these SNPs as diagnostic markers, we will subsequently estimate inversion frequencies in pooled resequencing *[Pool-Seq; 62]* data, where individuals with uncertain inversion status are pooled prior to DNA sequencing. In particular, we will utilize the DEST v.2.0 dataset *[63, 64]*, which is a collection of pooled whole-genome sequencing data from more than 700 *D. melanogaster* population samples, densely collected world-wide through time and space. Using the inversion-specific marker SNPs, we will estimate the inversion frequencies of our two focal inversions in the Pool-Seq data of each population sample and test how inversions influence genome-wide linkage disequilibrium and population structure. Furthermore, we will test for clinal patterns of the inversions in European and North American populations and investigate if these patterns can be explained by demography alone.

In summary, I will briefly introduce a broad variety of bioinformatic methods to explore the evolutionary effects of chromosomal inversions. The analyses that we will conduct will hopefully help to better understand the impact of chromosomal inversions on genome evolution and genetic variation in natural populations. By necessity, the pipeline described here, is quite specific to data available for the well-studied model organism *D. melanogaster*. However, I hope that the concepts, ideas and workflows described in the following pages prove helpful when designing similar analyses for other systems and research questions. The approach described here can, of course, also be used for other species, populations and inversion scenarios, provided that the necessary datasets are available.

## 2 Methods

### 2.1 Preparing the bioinformatics analyses pipeline

The full analysis pipeline of this book chapter can be found at https://github.com/capoony/Chapter6-InvPopGenomics. As a first step, all necessary software needs to be installed. This information can be found in a shell-script called dependencies.sh which is located in the shell/ folder. Here and throughout this chapter, the code blocks, as shown below, are highlighted by boxes with different fonts and colors according to the coding syntax of the BASH shell scripts, which is the main scripting language used in this analysis pipeline (besides specific scripts written in Python and R). It is possible to copy and paste the code snippets directly from this document and then execute them in a terminal window on a workstation computer or computer server with a LINUX operation system. However, I would strongly recommend to open the script main.sh, which is located in the shell/ folder and which contains the whole analysis pipeline shown here, in an integrated development environment (IDE) program such as the VScode (https://code.visualstudio.com/) editor and execute the individual commands from the main.sh script bit by bit.

```
### define working directory
WD=</Github/InvChapter> ## replace with path to the downloaded GitHub repo https://github.com/capoony/
Chapter6-InvPopGenomics

## install dependencies
sh ${WD}/shell/dependencies
```

After we make sure that all dependencies are correctly installed, we need to download sequencing data from the Short Read Archive (SRA: https://www.ncbi.nlm.nih.gov/sra) to obtain genomic information of individuals which have previously been characterized for their karyotype. These data will allow us to analyze the influence of inversions on genome-wide patterns of variation and differentiation. We will use the *Drosophila* Nexus dataset, a collection of more than one thousand whole-genome sequenced *D. melanogaster* individuals from all over the world ***[65, 66]***. We will focus on genomic data of individuals collected in Siavonga/Zambia with known karyotypes, which have been sequenced from haploid embryos ***[65, 66]*** by shotgun sequencing from paired-end libraries with Illumina technology. In a first step, we will download and process a metadata-table, which contains the sample names, the corresponding IDs from the SRA database and the inversion status of common inversions. We will use this information to select (up to) 20 individuals from each karyotype (INV and ST) for each of the two common cosmopolitan inversions, *In(2L)t* and *In(3R)Payne*. For each of the two inversions, we make sure that INV and ST individuals carry only the focal, but not the other inversion, in order to avoid possible linked effects on genomic variation if both inversions (or only the "wrong" inversion) are present. While such putative linked effects are indeed also interesting, their analysis is beyond the scope of this chapter. Here, we want to keep the analysis simple and focused on the individual effects of each inversion. After the samples that we will use for our downstream analyses have been selected from the metadata table with a specific R script based on the criteria described above, we will download their raw sequencing data from SRA. As you will see in the code block below, we will focus on the two inversions *In(2L)t* and *In(3R)Payne*, which we abbreviate with *IN2Lt* and *IN3RP*, respectively, for the sake of simplicity. The arrays *DATA*, *Chrom*, *Start* and *End* contain information of the genomic position for each of the two inversions.

```
## Get information of individual sequencing data and isolate samples with known inversion status
mkdir ${WD}/data
cd ${WD}/data

### download metadata Excel table for Drosophila Nexus dataset
wget http://johnpool.net/TableS1_individuals.xls

### process table and generate input files for downstream analyses, i.e., pick the ID's and SRA access
ion numbers for the first 20 individuals with inverted and standard karyotype, respectively.
Rscript ${WD}/scripts/ReadXLS.r ${WD}

### Define arrays with the inversion names, chromosome, start and end breakpoints; These data will be
reused in the whole pipeline for the sequential analysis and visualization of both focal inversions
DATA=("IN2Lt" "IN3RP")
Chrom=("2L" "3R")
Start=(2225744 16432209)
End=(13154180 24744010)

## Get read data from SRA
mkdir ${WD}/data/reads
mkdir ${WD}/shell/reads
conda activate sra-tools

### loop over both inversions
for index in ${!DATA[@]}; do
    INVERSION=${DATA[index]}

    ## read info from input file {WD}/data/${INVERSION}.txt that was generated above with ReadXLS.r
    while
        IFS=',' read -r ID SRR Inv
    do
        if [[ -f ${WD}/data/reads/${ID}_1.fastq.gz ]]; then
            continue
        fi

        echo """
        ## download reads and convert to FASTQ files
        fasterq-dump \
            --split-3 \
            -o ${ID} \
            -O ${WD}/data/reads \
            -e 8 \
            -f \
            -p \
            ${SRR}
        ## compress data
        gzip ${WD}/data/reads/${ID}*
        """ >${WD}/shell/reads/${ID}.sh
        sh ${WD}/shell/reads/${ID}.sh
    done <${WD}/data/${INVERSION}.txt
done
```

In the next step, we will first trim the raw sequencing data based on base-quality (PHRED score ≥ 18) and map the trimmed and filtered datasets against the *D. melanogaster* reference genome ***[v.6.57; 67]***, which we will download from FlyBase (https://flybase.org/). We will use a modified mapping pipeline from Kapun *et al.* ***[68]***, which further filters for PCR duplicates and improves the alignment of nucleotides around indels. In brief, this mapping pipeline first uses the program cutadapt to trim away sequencing adapter sequences which are still present in the raw reads and further trims flanking sequences if their base quality encoded as PHRED score is below 18. Since the sequencing data represent paired-end libraries, we will only retain intact read-pairs where each read has a minimum length of 25 bp for the downstream analyses. We use the program bwa-mem2 for mapping (i.e. aligning) the trimmed read-pairs against the reference genome before sorting the resulting BAM file, which we filter for mapping quality (PHRED >= 20) and correctly oriented read-pairs, by genomic position. Using the program Picard, we then identify and remove PCR duplicates in the dataset, which represents reads with identical mapping position that are particularly problematic in pooled sequencing data as they may bias allele frequency estimates. Finally, we use the program GATK to realign reads around indel positions to avoid false positive SNP variants due to misalignment. To keep the analysis pipeline concise, I collected all commands for the mapping pipeline in a separate shell script mapping.sh, which can be found in the shell/ folder.

```
## obtain D. melanogaster reference genome from FlyBase
cd ${WD}/data
wget -O dmel-6.57.fa.gz http://ftp.flybase.net/genomes/Drosophila_melanogaster/current/fasta/dmel-all-chromosome-r6.57.fasta.gz

## index the reference genome for the mapping pipeline
conda activate bwa-mem2
bwa-mem2 index dmel-6.57.fa.gz
gunzip -c dmel-6.57.fa.gz >dmel-6.57.fa
samtools faidx dmel-6.57.fa
samtools dict dmel-6.57.fa >dmel-6.57.dict
conda deactivate

## trim & map & sort & remove duplicates & realign around indels
for index in ${!DATA[@]}; do
    INVERSION=${DATA[index]}
    while
        IFS=',' read -r ID SRR Inv
    do
        ### run the mapping pipeline with 100 threads (modify to adjust to your system resources).
Note that this step may take quite some time
        sh ${WD}/shell/mapping.sh \
            ${WD}/data/reads/${ID}_1.fastq.gz \
            ${WD}/data/reads/${ID}_2.fastq.gz \
            ${ID} \
            ${WD}/mapping \
            ${WD}/data/dmel-6.57 \
            100 \
            ${WD}/scripts/gatk/GenomeAnalysisTK.jar
    done <${WD}/data/${INVERSION}.txt
done
```

Using the mapping pipeline, we aligned all reads against the *D. melanogaster* reference genome. Thus, we can now obtain the allelic information for each sample at every position in the reference genome, which is stored in the final BAM files. Since the sequencing data was generated from haploid embryos, we assume that there is only one correct allele present in each sample at a given genomic position. Please note that, in case only multiploid rather than haploid sequencing datasets are available for analyses, only individuals homozygous for either karyotype can be used for the bioinformatic approach described herein. We will now identify polymorphisms using the FreeBayes variant calling software and store the SNP information across all samples in VCF files. To speed these analyses up, I am using GNU parallel *[69]* with 100 threads, which splits the reference genome in chunks of 100,000 bp and process up to 100 chunks in parallel.

```bash
## SNP calling using freebayes with 100 threads
for index in ${!DATA[@]}; do

    INVERSION=${DATA[index]}
    while
        IFS=',' read -r ID SRR Inv
    do
        mkdir -p ${WD}/results/SNPs_${INVERSION}

        ### store the PATHs to all BAM files in a text, which will be used as the input for FreeBayes
        echo ${WD}/mapping/${ID}_RG.bam >>${WD}/mapping/BAMlist_${INVERSION}.txt

    done <${WD}/data/${INVERSION}.txt

    conda activate freebayes

    ### We assume ploidy = 1 and run FreeBayes in parallel by splitting the reference genome in chuncks
    of 100,000bps and use GNU parallel for multithreading. I am using 100 threads. Please adjust to your
    system.
    freebayes-parallel \
        <(fasta_generate_regions.py \
            ${WD}/data/dmel-6.57.fa.fai \
            100000) \
        100 \
        -f ${WD}/data/dmel-6.57.fa \
        -L ${WD}/mapping/BAMlist_${INVERSION}.txt \
        --ploidy 1 |
        gzip >${WD}/results/SNPs_${INVERSION}/SNPs_${INVERSION}.vcf.gz
    conda deactivate
done
```

## 2.2    Patterns of genomic variation and differentiation associated with different karyotpes in an African population from Zambia

In the next step, we will use population genetic statistics to compare genetic variation among Zambian INV and ST individuals with inverted and standard arrangement for the two inversions *In(2L)t* and *In(3R)Payne*, respectively. We will calculate Nei's $\pi$ *[nucleotide diversity; 70]* as an estimator of genetic variation in a population. Since VCFtools *[71]*, the program which we use to calculate the population genetic statistics, does not allow calculating these statistics from haploid VCF files, we will first convert the haploid VCF to a diploid version by duplicating the haploid haplotype for each individual. Then, we will calculate $\pi$ in 200,000 bp windows along the whole genome for the standard and the inverted individuals.

## 2.2.1   The influence of inversions on genetic variation

```
## calculate pi per karyotype
for index in ${!DATA[@]}; do
    INVERSION=${DATA[index]}

    mkdir ${WD}/data/${INVERSION}
    output_dir=${WD}/data/${INVERSION}

    ### split input file with sample IDs based on inversion status
    awk -F',' ' '
    {
        filename = $3 ".csv"
        filepath = "'$output_dir'/" filename
        if (filename == ".csv") next
        print $1 >> filepath
    }
    ' ${WD}/data/${INVERSION}.txt

    ### filter VCF for biallelic SNPs
    conda activate vcftools

    vcftools --gzvcf ${WD}/results/SNPs_${INVERSION}/SNPs_${INVERSION}.vcf.gz \
        --min-alleles 2 \
        --max-alleles 2 \
        --remove-indels \
        --recode \
        --out ${WD}/results/SNPs_${INVERSION}/SNPs_${INVERSION}

    gzip ${WD}/results/SNPs_${INVERSION}/SNPs_${INVERSION}.recode.vcf

    ### convert haploid VCF to diploid
    python ${WD}/scripts/hap2dip.py \
        --input ${WD}/results/SNPs_${INVERSION}/SNPs_${INVERSION}.recode.vcf.gz \
        --output ${WD}/results/SNPs_${INVERSION}/SNPs_${INVERSION}.recode_dip.vcf.gz

    for karyo in INV ST; do

        ### calculate pi in 200kbp windows
        vcftools --gzvcf ${WD}/results/SNPs_${INVERSION}/SNPs_${INVERSION}.recode_dip.vcf.gz \
            --keep ${WD}/data/${INVERSION}/${karyo}.csv \
            --window-pi 200000 \
            --out ${WD}/results/SNPs_${INVERSION}/${INVERSION}_${karyo}_pi
    done

    ## combine pi of INV and ST chromosomes
    awk 'NR ==1 {print $0"\tType"}' ${WD}/results/SNPs_${INVERSION}/${INVERSION}_INV_pi.windowed.pi >$
{WD}/results/SNPs_${INVERSION}/${INVERSION}_pi.tsv
    awk 'NR>1  {print $0"\tINV"}' ${WD}/results/SNPs_${INVERSION}/${INVERSION}_INV_pi.windowed.pi >>${
WD}/results/SNPs_${INVERSION}/${INVERSION}_pi.tsv
    awk 'NR>1  {print $0"\tST"}' ${WD}/results/SNPs_${INVERSION}/${INVERSION}_ST_pi.windowed.pi >>${WD
}/results/SNPs_${INVERSION}/${INVERSION}_pi.tsv

done
```

After we merged all $\pi$ estimates for both karyotypes in a single file for each inversion, we plot the genome-wide patterns in line-plots for each chromosome and each karyotype. Here, as well as in the rest of this pipeline, we will use ggplot from the tidyverse package (72) in R (73) for creating graphics. For the sake of brevity, I collected all R code in separate R-scripts, which can all be found in the scripts/ folder.

```
### plot PI as line plot
for index in ${!DATA[@]}; do

    INVERSION=${DATA[index]}
    St=${Start[index]}
    En=${End[index]}
    Ch=${Chrom[index]}

    Rscript ${WD}/scripts/Plot_pi.r \
        ${INVERSION} \
        ${Ch} \
        ${St} \
        ${En} \
        ${WD}

done
```

With the exception of the genomic regions of the two inversions (*In(2L)t*: Figure 2A; and *In(3R)Payne*: Fig. 2B), we find that $\pi$-values are very similar, irrespective of the karyotype (INV: red and ST: blue), across the whole genome. In contrast, we see that genetic variation within the inverted regions is markedly reduced in INV individuals, which suggests that the population sizes of both inversions are smaller their standard arrangements. These patterns may, moreover, also indicate that gene flux and novel mutations did not have enough time yet to reconstitute genetic variation similar to the ancestral arrangement *[22]*. In addition, we see that the reduction of genetic variation is not equal across the whole inverted region. Rather, the reduction is strongest close to the breakpoints. This indicates that gene flux in the center of the inversion has, at least partially, shifted genetic variation from the standard arrangement into the inverted chromosomes. In the case of *In(2L)t*, we further find that the effect of the inversion on genetic variation is not confined to the region spanned by the inversion but also spreads millions of base pairs beyond the breakpoints (Figure 2A). Finally, we also observe that the distribution of nucleotide diversity varies along each chromosome and is strongly reduced close to the centro- and telomers (Figure 2). These particular regions are characterized by reduced recombination rates, which influences the extent of background selection, leading to reduced variation in these chromosomal regions *[74]*.

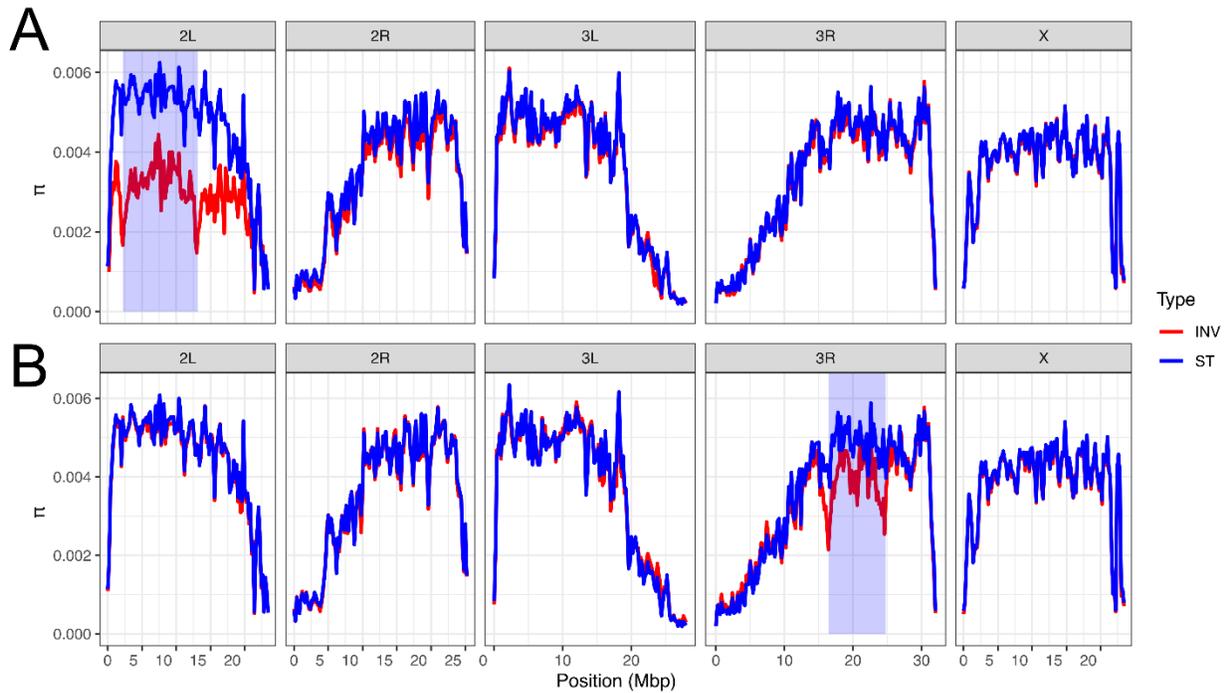

**Figure 2**: Line plots showing the distribution of nucleotide diversity $\pi$ along the genome averaged in 200kbp non-overlapping windows for inverted (red) and non-inverted (blue individuals). The genomic regions spanned by the inversions *In(2L)t* (Panel A) and *In(3R)Payne* (Panel B) are highlighted by transparent blue boxes.

### 2.2.2   The influence of inversions on genetic differentiation

In the next part, we will use the diploid SNP dataset generated above to calculate $F_{ST}$ estimates among the INV and ST individuals for each inversion using the method of Weir & Cockerham *[75]* as implemented in VCFtools. The fixation index $F_{ST}$ summarizes genetic structure and is scaled between zero (no differentiation) and one (complete differentiation). We will use this metric to identify single SNPs, which are strongly differentiated between the karyotypes. In addition, we will calculate $F_{ST}$ averaged in 200,000 bp windows to find genomic regions, where many neighboring SNPs show similar differentiation patterns.

```
## calculate F_ST between karyotypes
for index in ${!DATA[@]}; do
    INVERSION=${DATA[index]}

    conda activate vcftools

    ## calculate F_ST per SNP
    vcftools --gzvcf ${WD}/results/SNPs_${INVERSION}/SNPs_${INVERSION}.recode_dip.vcf.gz \
        --weir-fst-pop ${WD}/data/${INVERSION}/INV.csv \
        --weir-fst-pop ${WD}/data/${INVERSION}/ST.csv \
        --out ${WD}/results/SNPs_${INVERSION}/${INVERSION}.fst

    ## calculate F_ST in 200kbp windows
    vcftools --gzvcf ${WD}/results/SNPs_${INVERSION}/SNPs_${INVERSION}.recode_dip.vcf.gz \
        --weir-fst-pop ${WD}/data/${INVERSION}/INV.csv \
        --weir-fst-pop ${WD}/data/${INVERSION}/ST.csv \
        --fst-window-size 200000 \
        --out ${WD}/results/SNPs_${INVERSION}/${INVERSION}_window.fst

    conda deactivate
done
```

We then plot both SNP-wise $F_{ST}$ as well as $F_{ST}$ values averaged in 200kbp windows. These types of plots are so-called Manhattan plots, where each dot represents a polymorphic genomic position along the x-axis and the corresponding $F_{ST}$ value are shown on the y-axis. On top, we are plotting the window-wise $F_{ST}$ as a line and highlight the region of the corresponding inversion by a transparent blue box.

```
for index in ${!DATA[@]}; do

    INVERSION=${DATA[index]}
    St=${Start[index]}
    En=${End[index]}
    Ch=${Chrom[index]}

    ### plot FST as Manhattan Plots
    Rscript ${WD}/scripts/Plot_fst.r \
        ${INVERSION} \
        ${Ch} \
        ${St} \
        ${En} \
        ${WD}
done
```

As you can see in Figure 3 for *In(2l)t* (A) and *In(3R)Payne* (B), genetic differentiation is elevated among the karyotypes within the inversion and particularly at and around the inversion breakpoints. These patterns suggest that novel mutations building up over time in the proximity of the inversion breakpoints result in strong differentiation. Consistent with theory, the suppression of recombination prevents genetic homogenization among the karyotypes across the whole inverted region, but specifically at the breakpoints *[22]*. Particularly for *In(3R)Payne*, we observe a typical "suspension bridge" pattern of genetic differentiation within the inversion *[29]*, which suggests that gene flux in distance to the breakpoints and towards the center of the inversion has led to a homogenization of genetic variation across the two karyotypes *[27]*. Similar to what we observed in Figure 2, we also observe that patterns of differentiation spread way beyond the inversion breakpoints, as shown for *In(2L)t*, which further emphasizes the genome-wide impact of inversions on genetic variation.

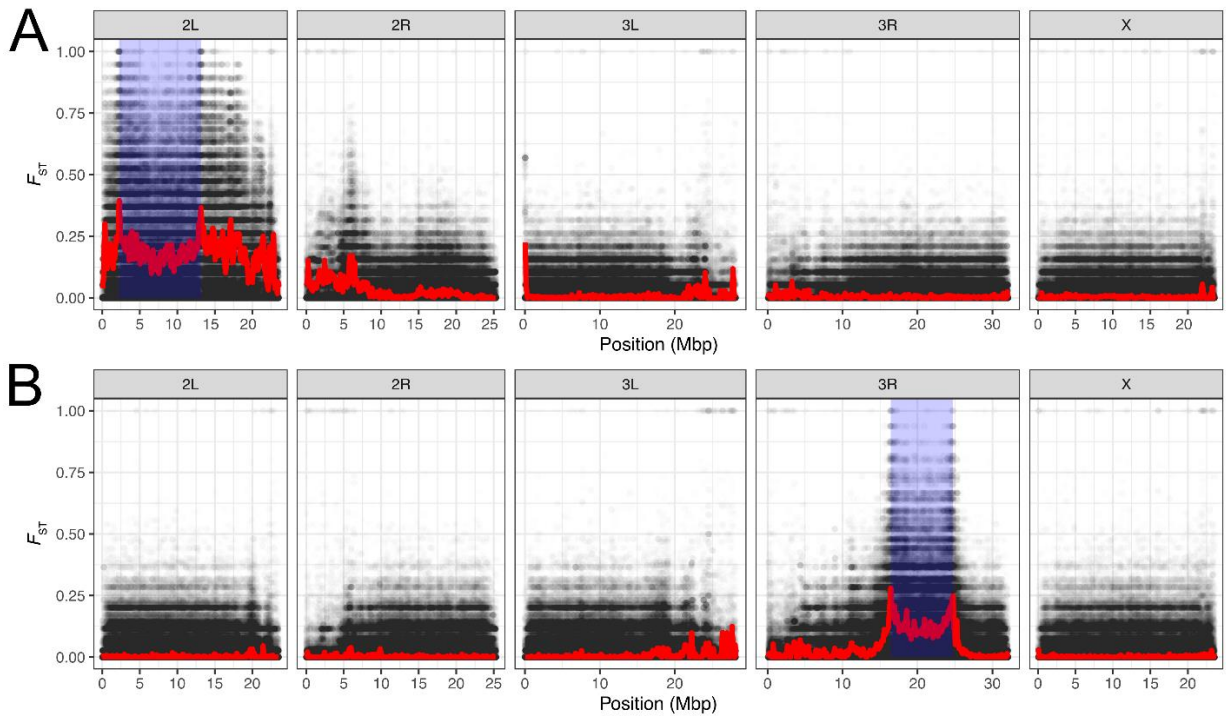

**Figure 3**: Manhattan plots showing the distribution of SNP-wise $F_{ST}$ between INV and ST individuals along the genome. In addition, averaged $F_{ST}$ values in 200kbp windows are overlaid as line plots in red and the genomic regions spanned by the inversions *In(2L)t* (A) and *In(3R)Payne* (B) are highlighted by transparent blue boxes.

### 2.3    SNPs in strong linkage disequilibrium with different karyotypes

Several SNPs that are clustered at the inversion breakpoints in the Manhattan plots of Figure 3 show very high $F_{ST}$ values, which indicates complete or near complete fixation for different alleles between the two karyotypes. We therefore assume that many SNPs are in complete linkage disequilibrium (LD) with the inversion - at least in the specific Zambian population sample that we are analyzing. In those SNPs, one allele is associated with the inverted karyotype (INV) and the other with the standard arrangement (ST). This makes it possible to use these SNPs as diagnostic markers that allow to test if the sequencing data of given individual with unknown karyotype is carrying the inversion simply by tracing for the inversion-specific allele at the corresponding diagnostic markers. Furthermore, it is possible to estimate the frequency of inverted chromosomes in pooled sequencing (Pool-Seq) data, where multiple individuals are pooled prior to DNA extraction and the pool of DNA is then sequenced jointly *[61]*. In this kind of datasets, it is assumed that the allele frequency in the pooled sequencing reliably estimates the actual frequency of the allele in the population from which the pooled individuals were randomly sampled *[76]*. Thus, the median frequency of the inversion-specific alleles in the pooled dataset should roughly correspond to the inversion frequency given that these SNPs are in tight LD with the inversion. However, I need to caution here, that these markers should - at best - be applied only to sequencing data from samples collected in the same broader geographic region, or that diagnostic marker SNPs are defined using a mixed sample of individuals with known karyotype from all areas where the corresponding inversion occurs. The evolutionary history of inversions with a broad geographic distribution may in fact be very complex and characterized by the emergence and fixation of different SNPs within the inversion in different geographic regions *[31]*.

2.3.1 Inversion-specific diagnostic marker SNPs

In the following, we will use a custom script that searches the VCF files with samples specific to the *In(2L)t* and *In(3R)Payne* analyses for potential candidate SNPs. Specifically, we will focus on SNPs that are in full LD with either of the two focal inversions and which are located within 200kbp distance to an inversion breakpoint. For these SNPs, we will then isolate the alleles that are fixed within the inverted chromosomes.

```
## obtain diagnostic SNPs for each inversion
for index in ${!DATA[@]}; do

    INVERSION=${DATA[index]}
    St=${Start[index]}
    En=${End[index]}
    Ch=${Chrom[index]}

    ### store the chromosome, start and endpoints of each inversion as a comma-separated string
    BP="${Ch},${St},${En}"

    ### only retain the header and the rows on the inversion-specific chromosome and focus on the focal
    individuals that are either INV or ST
    gunzip -c ${WD}/results/SNPs_${INVERSION}/SNPs_${INVERSION}.recode.vcf.gz |
        awk -v Ch=${Ch} '$1~/^#/|| $1 == Ch' |
        python ${WD}/scripts/DiagnosticSNPs.py \
            --range 200000 \
            --breakpoints ${BP} \
            --input - \
            --output ${WD}/results/SNPs_${INVERSION}/${INVERSION} \
            --MinCov 10 \
            --Variant ${WD}/data/${INVERSION}.txt
done
```

Our analysis results in 62 and 26 diagnostic SNPs for *In(2L)t* and *In(3R)Payne*, respectively. In the following paragraphs, we will use these marker SNPs to indirectly infer inversion frequencies in other genomic datasets, but before that, we will test if inversions influence population structure in *D. melanogaster* population samples from North America and Europe without prior information on inversion frequencies in the corresponding samples.

2.3.2   The influence of inversions on population structure

To this end, we will use the largest Pool-Seq dataset of natural *D. melanogaster* populations available to date. The DEST v.2.0 dataset combines more than 700 world-wide population samples of vinegar flies from different sources that were densely collected through space and time, yet mostly from North American and from European populations *[63, 64, 68]*. All shotgun sequence data were processed with a standardized trimming and mapping pipeline (as described above) prior to joint SNP calling with the heuristic variant caller PoolSNP *[68]*. Moreover, DEST v.2.0. also provides rich metadata, including detailed information on the sampling date and location, basic sequencing statistics (such as read depths, SNP counts, etc.) and recommendations based on data quality assessments *[64]*. With the help of these metadata, we will subset the full data and only consider population samples of high quality and focus on samples collected from North America and Europe, respectively. Then, we will apply principal component analyses (PCA) to the allele frequency data to identify genome-wide differences among the samples in our datasets. Essentially, the first few orthogonal PC-axes capture most of the genetic variation shared by genome-wide SNPs and reflect the shared evolutionary history of the populations in the dataset *[77]*. For these reasons, PCA is a very popular model-free method to quantify population structure. In our example, we will compare the results of PCA applied to SNPs located either within the genomic region spanned by each of the inversions or away from these inversion-specific genomic regions. These analyses will reveal if the inversions have an influence on population structure in their genomic region and how these patterns differ from genome-wide estimates.

As a first step, we will download both the DEST v.2.0 SNP data in VCF file-format and the corresponding metadata as a comma-separated (CSV) table from the DEST website. In addition, we will download two scripts from the DEST pipeline that are needed for the downstream analyses.

```
### download VCF file and metadata for DEST dataset
cd ${WD}/data
wget -O DEST.vcf.gz http://berglandlab.uvadcos.io/vcf/dest.all.PoolSNP.001.50.3May2024.ann.vcf.gz
wget -O meta.csv https://raw.githubusercontent.com/DEST-bio/DESTv2/main/populationInfo/dest_v2.samps_3May2024.csv

###  download sripts
cd ${WD}/scripts
wget https://raw.githubusercontent.com/DEST-bio/DESTv2_data_paper/main/16.Inversions/scripts/VCF2sync.py
wget https://raw.githubusercontent.com/DEST-bio/DESTv2_data_paper/main/16.Inversions/scripts/overlap_in_SNPs.py
```

Using the metadata table, we will identify all samples that we are going to include in our continent-wide analyses of populations from North America and Europe. Furthermore, we will apply several filtering steps to the VCF file and, based on metadata information, we will construct two continent-specific datasets consisting of allele frequency data that we will use for all downstream analyses. Specifically, we will (1) isolate continent-specific populations, (2) remove problematic samples (based on DEST recommendations), remove (3) populations with < 15-fold average read depth, (4) only retain biallelic SNPs, (5) convert the allele counts to frequencies of the reference allele and obtain (6) read-depths for each position and population sample. Finally, (7) we will restrict our analyses to 50,000 randomly drawn genome-wide SNPs. The final files will represent a two-dimensional matrix of reference allele frequencies (based on the reference genome), where rows represent polymorphic genomic positions and columns represent population samples.

```
### Split metadata by continent

## remove single quotes from metadata table
sed -i "s/'//g" ${WD}/data/meta.csv

## split by continent
awk -F "," '$6 =="Europe" {print $1}' ${WD}/data/meta.csv >${WD}/data/Europe.ids
awk -F "," '$6 =="North_America" {print $1}' ${WD}/data/meta.csv >${WD}/data/NorthAmerica.ids

## get data for populations that did not pass the quality criteria (no PASS and average read depths <
15)
awk -F "," '$(NF-7) !="Pass" || $(NF-9)<15 {print $1"\t"$(NF-7)"\t"$(NF-9)}' ${WD}/data/meta.csv
>${WD}/data/REMOVE.ids

### subset the VCF file

mkdir ${WD}/results/SNPs

for continent in NorthAmerica Europe; do

    conda activate vcftools

    ## decompress VCF file
    pigz -dc ${WD}/data/DEST.vcf.gz |

        ## keep header and position with only one alternative allele
        awk '$0~/^\#/ || length($5)==1' |

        ## keep continental data and remove bad quality samples
        vcftools --vcf - \
            --keep ${WD}/data/${continent}.ids \
            --remove ${WD}/data/REMOVE.ids \
            --recode \
            --stdout |

        ## remove rows with missing data
        grep -v "\./\." |

        ## randomly sample 50,000 SNPs
        python ${WD}/scripts/SubsampleVCF.py \
            --input - \
            --snps 50000 |

        ## convert VCF to allele frequencies and weights (of the reference allele)
        python ${WD}/scripts/vcf2af.py \
            --input - \
            --output ${WD}/results/SNPs/${continent}

done
```

Now, we will employ separate principal component analyses (PCA) for European and North American samples to test if the genetic variation in the genomic region spanned by an inversion influences signals of population structure. To this end, we will execute the R script PCA_inv.r in the scripts/ folder to carry out the following analysis steps: at first, we will load the allele frequency datasets generated above and split the data in two subsets, where one subset contains SNPs located within the genomic region of a given inversion and the other contains SNPs from the remaining (inversion-free) part of the genome. Then, we will perform PCA on the transposed allele frequency matrices, where columns represent chromosomal positions and rows represent population samples. Finally, we will use metadata information to highlight the country/county of origin in scatterplots that show the first two PC-axes for each subset per continent and inversion.

```
### use PCA to test for patterns inside and outside the genomic region spanned by an inversion
Rscript ${WD}/scripts/PCA_Inv.r \
    ${WD}
```

The scatterplots in Figure 4 show the first two PC-axes, which together explain between 8%-12% of the total genetic variation of all SNPs included in the analysis. Notably, the PC-scores of genome-wide SNPs in Europe (A) and North America (B) cluster populations mostly according to expectations based on geography. Conversely, PC-scores calculated from SNPs inside the breakpoints of *In(2L)t* and *In(3R)Payne* are much more compressed and appear to mostly follow some diagonals rather than clustering according to geography - particularly for PC1 *[see also 78]*. We may thus speculate that genetic variation associated with inversion, and thus inversion frequencies, strongly contribute to the observed patterns in the investigated populations.

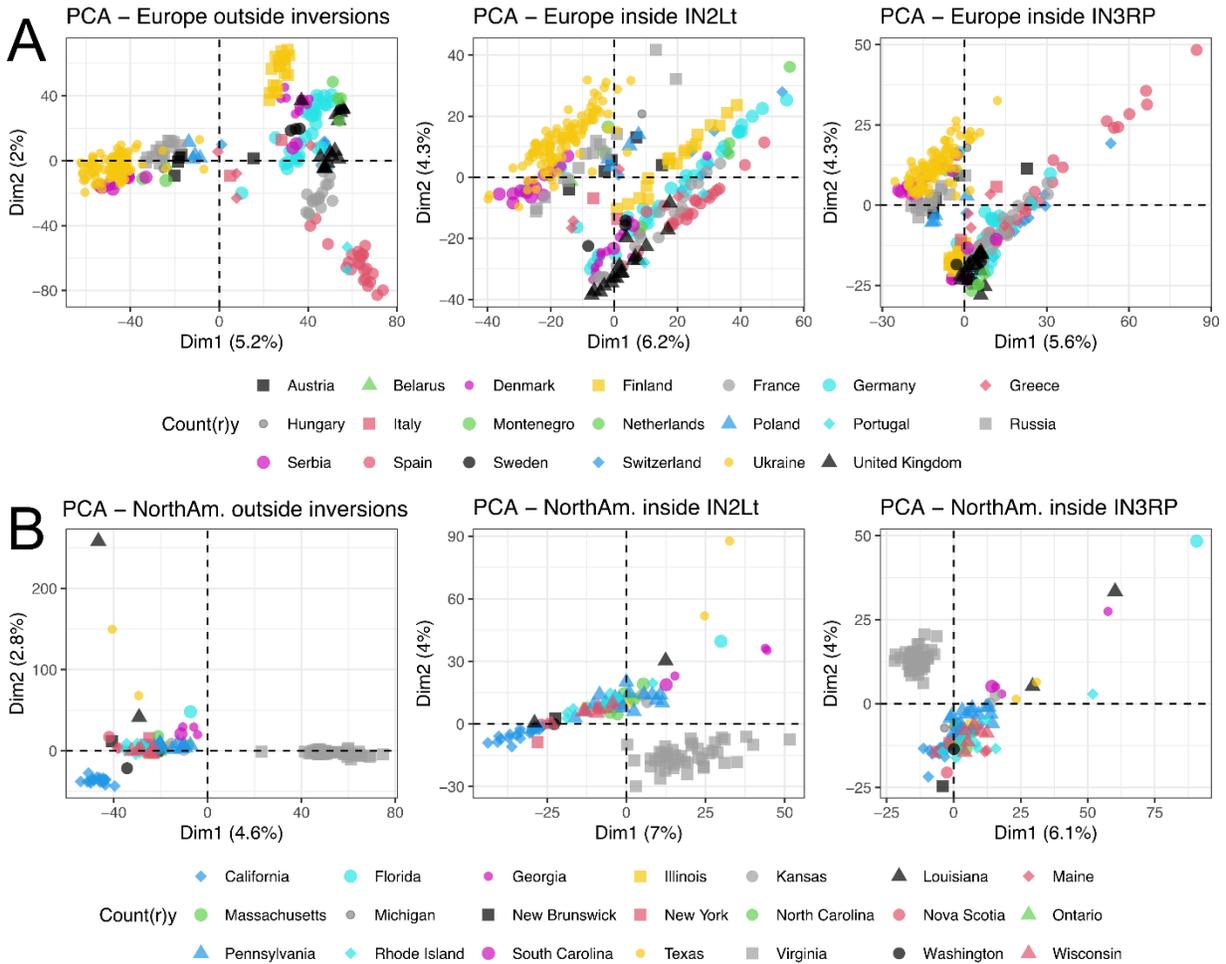

## 2.4 Inversion frequencies in Pool-Seq data

We will now take advantage of the diagnostic marker SNPs that we isolated above and will estimate inversion frequencies in each of the pooled population samples in Europe and North America, respectively. This will allow us to test more directly how inversions influence genetic variation and population structure and if the two inversions exhibit clinal variation.

### 2.4.1 Estimating inversion frequencies in Pool-Seq data with diagnostic markers

We will now convert the VCF file to the SYNC file format using the Python script VCF2sync.py from the DEST pipeline. SYNC files are commonly used to store allele counts in pooled sequencing data as colon-separated lists in the form "A:T:C:G:N:Del" for each population sample and position. We will then obtain allele counts from the SYNC file at the positions of inversion-specific marker SNPs that are present in the DEST dataset using the Python script overlap_in_SNPs.py. To speed these calculations up, I am using GNU parallel with 100 threads.

```
### convert VCF to SYNC file format
conda activate parallel
gunzip -c ${WD}/data/DEST.vcf.gz |
    parallel \
        --jobs 100 \
        --pipe \
        -k \
        --cat python3 ${WD}/scripts/VCF2sync.py \
        --input {} |
    gzip >${WD}/data/DEST.sync.gz

### Get positions at inversion specific marker SNPs
for index in ${!DATA[@]}; do

    INVERSION=${DATA[index]}
    gunzip -c ${WD}/data/DEST.sync.gz |
        parallel \
            --pipe \
            --jobs 100 \
            -k \
            --cat python3 ${WD}/scripts/overlap_in_SNPs.py \
            --source ${WD}/results/SNPs_${INVERSION}/${INVERSION}_diag.txt \
            --target {} \
            >${WD}/data/DEST_${INVERSION}.sync
done
```

For each population and for each of the two inversions, we now calculate the median frequency of the inversion-specific alleles across all diagnostic markers to obtain an estimate of the corresponding inversion frequency with a custom Python script. Before that, we need to obtain the names of all samples in the VCF file in the correct order and then output the estimated inversion frequencies as a tab-delimited file.

```bash
### get the names of all samples in the VCF file and store as an array
NAMES=$(gunzip -c ${WD}/data/DEST.vcf.gz | head -150 | awk '/^#C/' | cut -f10- | tr '\t' ',')

# Calculate median frequencies for marker SNPs
for index in ${!DATA[@]}; do

    INVERSION=${DATA[index]}

    python3 ${WD}/scripts/inversion_freqs.py \
        --marker ${WD}/results/SNPs_${INVERSION}/${INVERSION}_diag.txt \
        --input ${WD}/data/DEST_${INVERSION}.sync \
        --names $NAMES \
        --inv ${INVERSION} \
        >${WD}/results/SNPs_${INVERSION}/${INVERSION}.af

done
```

To visually inspect the distribution of inversion-specific alleles across all diagnostic markers for each inversion and population sample, we plot histograms of all inversion-specific allele frequencies and the actual allele frequencies of all diagnostic SNPs against their genomic position. In addition, we plot the median frequency, which we consider the estimated inversion frequency, as a dashed line atop the frequency histogram. We therefore need to first generate a table with the inversion-specific allele frequencies of the diagnostic SNPs for all population samples in the DEST v.2.0 VCF file. Then, we generate the above-mentioned plots of allele frequencies in R using the script Plot_InvMarker.r in the scripts/ folder.

```bash
### generate plots for each population
for index in ${!DATA[@]}; do

    INVERSION=${DATA[index]}
    Ch=${Chrom[index]}

    ### convert VCF to allele frequency table for each SNP and population sample
    gunzip -c ${WD}/data/DEST.vcf.gz |
        awk -v Ch=${Ch} '$1~/^#/|| $1 == Ch' |
        python3 ${WD}/scripts/AFbyAllele.py \
            --input - \
            --diag ${WD}/results/SNPs_${INVERSION}/${INVERSION}_diag.txt \
            >${WD}/results/SNPs_${INVERSION}/${INVERSION}_pos.af

    ### make plots in R
    Rscript ${WD}/scripts/Plot_InvMarker.r \
        ${INVERSION} \
        ${WD}
done
```

The example in Figure 5 shows the distribution of inversion-specific alleles for *In(2L)t* in a population sample collected in 2015 close to Mautern, in the beautiful Wachau area along the Danube river in Austria. As you can see in the scatterplot, two sets of SNPs are located around the breakpoints of *In(2L)t* and the frequencies of the inversion-specific alleles span from 0% to more than 60%, which is quite a broad range and presumably the result of sampling error in the Pool-Seq data. Assume that even if a diagnostic SNP is in full linkage disequilibrium with the inversion, it will not necessarily depict the "true" frequency of the inversion in the population due to binomial sampling. When sequencing the pooled DNA from multiple samples with NGS methods such as Illumina, we are usually sampling (i.e., sequencing) 50 to 100 DNA fragments at every genomic position, which corresponds to a 50 to 100-fold sequencing depth. This is, of course, only a very small fraction of the millions of copies of genomic DNA in the extracted DNA. The resulting sampling error leads to deviations from the expected frequency (i.e., the true allele frequency in all the DNA copies) and these deviations become even larger when the sequencing depths are lower. However, if we further assume that each SNP is a reliable estimator of the "true" inversion frequency due to perfect linkage disequilibrium, we expect that the inferred frequencies across all marker SNPs roughly follow a binomial distribution, where the sequencing depth corresponds to the number of trials *n* and the expected inversion frequency corresponds to the number of successes *k*. However, other factors, such as sequencing and mapping errors or imperfect LD of some diagnostic SNPs in certain geographic regions (see also above in the introduction to paragraph 3) may also influence the distribution of frequencies. Rather than calculating the mean frequencies across all markers, we use the median to estimate the population inversion frequency and compare inversion patterns in all population samples, since this statistic is more robust to asymmetric distributions. In our example in Figure 5, the median is shown as a dashed red line at approximately 25% frequency.

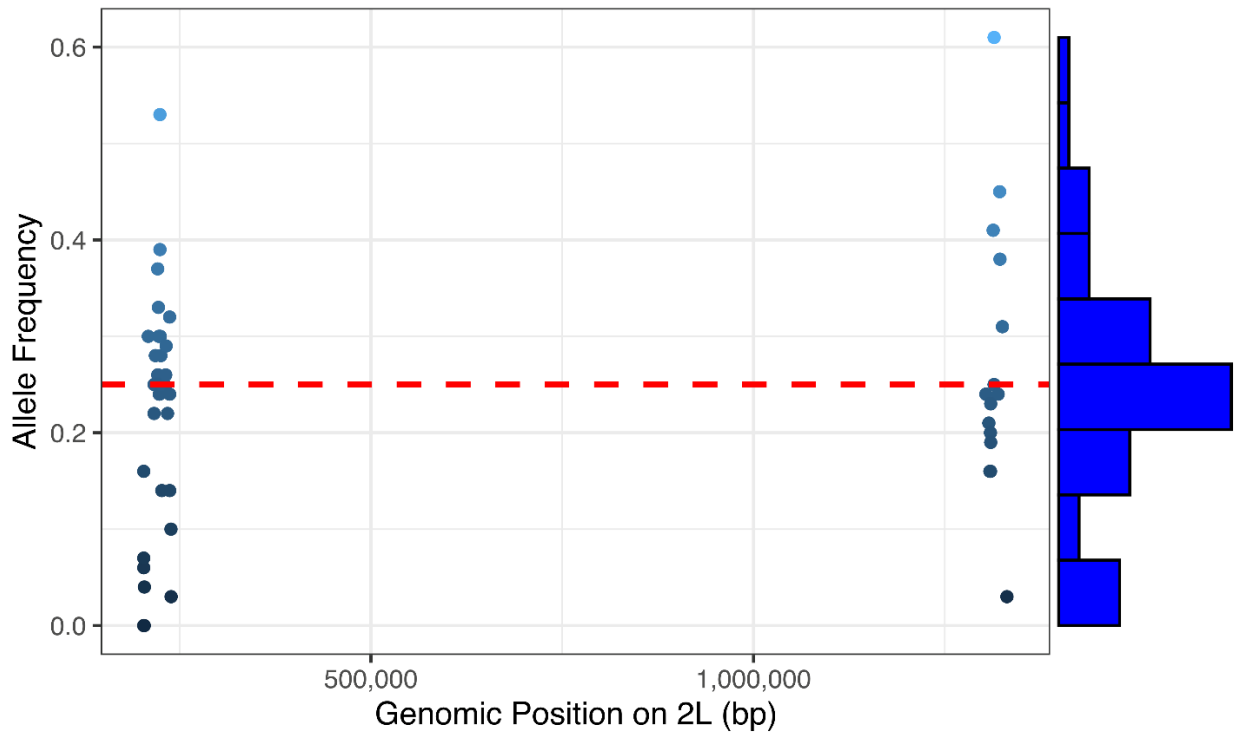

**Figure 5**: Frequencies of alleles from the diagnostic marker SNPs for *In(2Lt)* that are in strong LD plotted against their genomic position on chromosome 2L for one population sample collected in 2015 in Mautern, Lower Austria. A histogram of allele frequencies plotted on the right along the y-axis shows the range of frequencies across all marker SNPs. In addition, we plot the median allele frequency across all markers as a horizontal dashed line in red, which we consider the estimated inversion frequency in the population sample.

<u>2.4.2   The influence of inversions on population structure revisited</u>

In paragraph 3.2 we found that the genomic regions spanned by inversions differ in patterns of population structure from the remaining genome. We speculated that inversion frequencies may play an important role. Now that we have estimated inversion frequencies in all populations based on the diagnostic marker SNPs, we can test if this hypothesis is true. We will test in particular for correlations between the inversion frequencies and the scores of PC1 based on SNPs located either inside and outside the genomic region spanned by each inversion.

We will execute the R-script Plot_PCAInvFreq.r, which will create scatterplots based on the PC-scores of the first PC-axis (Dim.1) and the inversion frequency for *In(2L)t* and *In(3R)Payne*, fit a linear regression line to each of the plots and add the adjusted $R^2$ value (the determinant of correlation) in the top-right corner of each plot, which describes the proportion of the variance explained by the correlation.

```
## does the Inv Frequency influence the PCA results?
for index in ${!DATA[@]}; do

    INVERSION=${DATA[index]}
    Rscript ${WD}/scripts/Plot_PCAInvFreq.r \
        ${INVERSION} \
        ${WD}
done
```

Consistent with our hypothesis, we can see in Figure 6 that all plots on the left side, which show scatterplots of inversion frequencies and PC1, are characterized by very strong and highly significant correlations, which explain 54% to 81% of the total variance. Conversely, we see that the inversion frequency barely explains any variance in the population structure inferred from genome-wide SNPs outside the inversions. We therefore conclude that our initial hypothesis was correct and that the two inversions indeed have a major impact on the distribution of genetic variation and signals of population structure. Thus, not accounting for chromosomal inversions when testing for population structure can lead to biased conclusions.

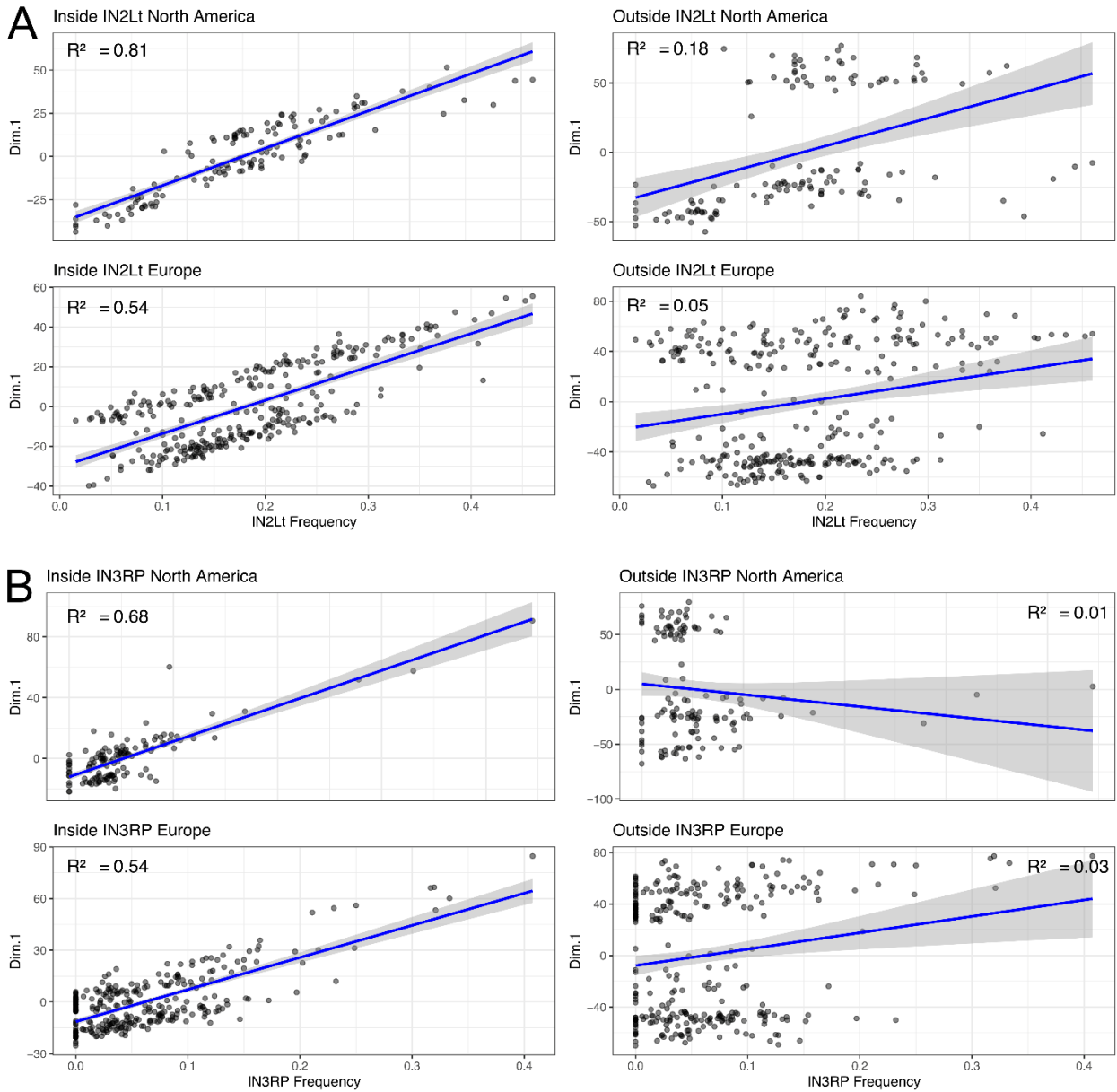

**Figure 6**: Scatterplots showing the association between inversion frequencies and the first PC-axis based on SNPs either located inside (left panels) our outside (right panels) the genomic region spanned by *In(2L)t* (A) or *In(3R)Payne* (B) for Europe (rows 2 and 4) and North America (rows 1 and 3). Regression lines based on linear regression are shown in blue and the determinants of correlation (adjusted *R²* value) are printed in the top-right corner of each plot.

### 2.4.3   The influence of inversions on genetic differentiation revisited

In the next analysis step, we will extend the analysis detailed in paragraph 2.3.2, where we investigated the influence of inversions on genome-wide differentiation. While these analyses were restricted to a single population in Zambia, we will now investigate hundreds of populations in Europe and North America. Previously, we used $F_{ST}$ to quantify (linked) differences between the two karyotypes, but now, we will use logistic regression to test, for each SNP in the genome, whether its frequency is significantly correlated with the inversion frequency for either *In(2L)t* or *In(3R)Payne* across all populations of the data sets of the two continents. We assume that SNPs that are in strong LD, either due to physical proximity to the inversion or due to co-evolution, (i.e., statistical linkage) will be characterized by a statistically significant regression model.

Again, the actual analysis steps are stored in the R-script PlotInvLD.r, where the allele frequency matrices and the corresponding read depths are loaded for each continent. Then, logistic regressions will be calculated for each SNP, where the allele frequencies of a given SNP is the dependent variable, the inversion frequency is the independent variable, and the read depth is the weight. For each inversion and continent, we then generate Manhattan plots, where the x-axis shows the genomic position of a given SNP and the y-axis shows the negative $\log_{10}$-transformed *p*-value of a likelihood ratio test of the logistic regression.

```
## calculate SNP-wise logistic regressions testing for associations between SNP allele frequencies and
inversion frequencies to test for linkage between SNPs and inversions for Europe and North America

for index in ${!DATA[@]}; do

    INVERSION=${DATA[index]}
    St=${Start[index]}
    En=${End[index]}
    Ch=${Chrom[index]}

    Rscript ${WD}/scripts/PlotInvLD.r \
        ${INVERSION} \
        ${Ch} \
        ${St} \
        ${En} \
        ${WD}

done
```

All four Manhattan plots shown in Figure 7 reveal that SNPs whose allele frequencies are influenced by the inversion frequencies (in hundreds of natural populations both in Europe and in North America), are strongly enriched within the genomic regions covered by the corresponding inversions, as indicated by elevated -$\log_{10}$-$p$-values. Similar to the results from paragraph 2.2 we also find that the LD is not only restricted to the genomic regions within the inversion breakpoints. Particularly for *In(2L)t*, we again find that many SNPs downstream of the distal breakpoint also show highly significant correlations with the inversion, which indicates a suppression of recombination across large parts of this chromosomal arm ***[56]***. Since this analysis only includes 50,000 genome-wide polymorphisms, we do not see a strong enrichment of SNPs in high LD at the breakpoints.

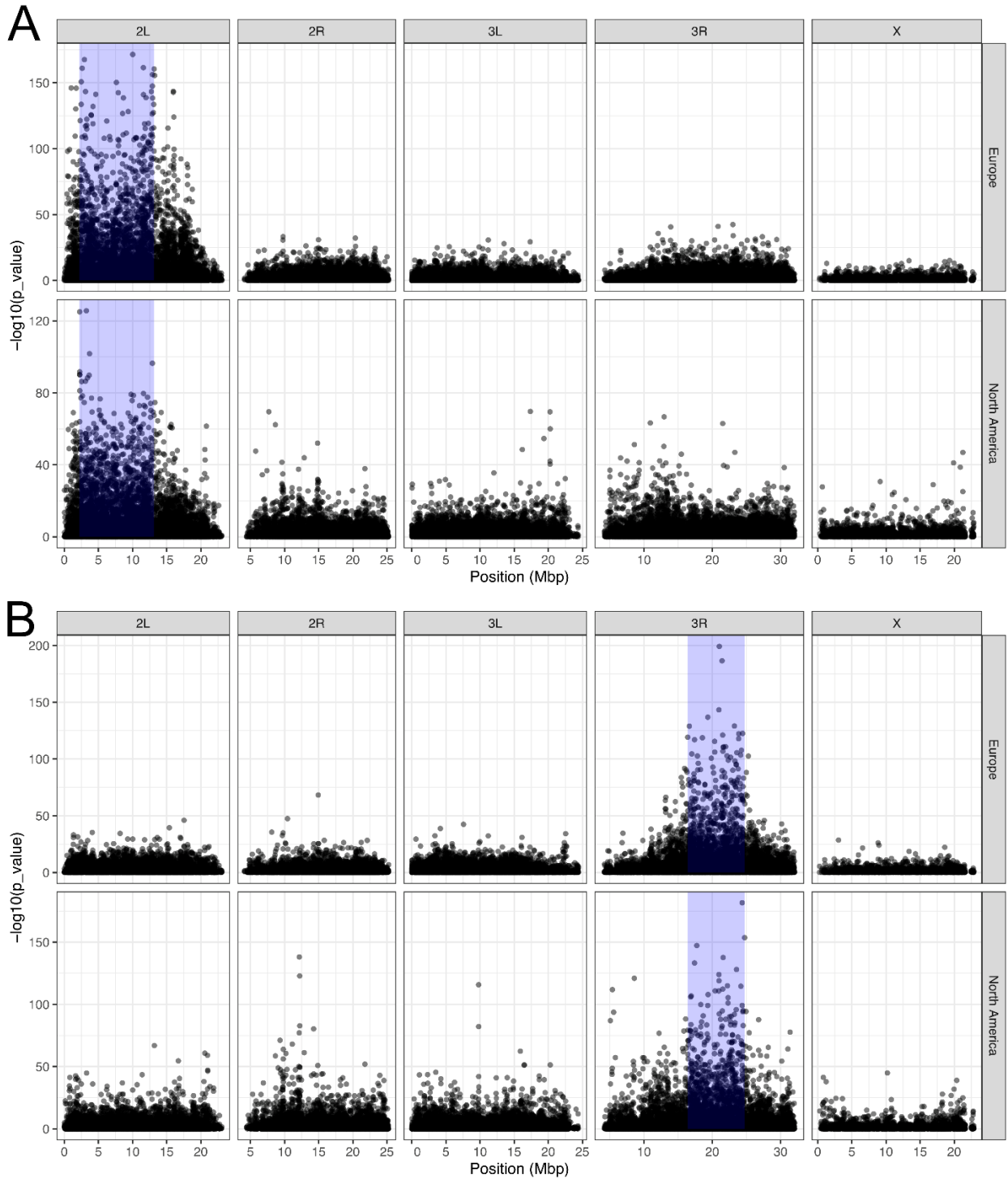

**Figure 7**: Manhattan plots showing the results of SNP-wise logistic regressions that test for associations between inversion frequencies and allele frequencies at a given SNP. The *p*-values are -log₁₀ - transformed and plotted along

the y-axis. In addition, we highlight the genomic position of *In(2L)t* (A) and *In(3R)Payne* (B) by blue transparent boxes in Europe (rows 1 and 3) and North America (rows 2 and 4).

## 2.4.4   Clinal patterns of chromosomal inversions

A large body of literature has documented clinal distributions, i.e., gradual frequency changes of several inversions along environmental gradients. One particularly prominent example is the latitudinal cline of *In(3R)Payne* along the North American East Coast *[35, 56]* and in Australia *[79]*. Here, we will test if the inversion frequencies estimated by our diagnostic marker SNPs show these expected patterns along latitudinal and longitudinal gradients in Europe and North America. Moreover, we will test if these patterns can be explained by neutral evolution alone (for example, by mechanisms such as isolation by distance, secondary contact, admixture *[80, 81]*, etc.). In contrast, if the inversion has evolved a clinal distribution due to spatially varying selection, we would assume that the clinal patterns strongly deviate from the genomic background. However, sometimes it is difficult to distinguish neutral and adaptive genomic signals. If, for example, population samples show very strong population structure along an environmental gradient, it may be misleading to only investigate allele frequency differences at a single gene (or inversion) that is considered a candidate for selection. A signal for strong association of allele frequencies at the gene (or inversion) with the environmental gradient may be strongly confounded by (unknown) population structure, which would lead to similar patterns of genetic differentiation between the populations across the whole genome *[81, 82]*. Thus, a signal of differentiation at a single locus may be misinterpreted as the result of selection, while it is in reality the result of the unknown genome-wide evolutionary history. To account for this, we will employ a statistical approach from landscape genomics *[83]*. Latent factor mixed models *[LFMM; 84, 85]* first identify genome-wide patterns of genetic variation by PCA (very similar to our approach in 2.3.2) assuming that the first few PC axes capture genome-wide differences which are predominantly the result of the demographic history of the sampled populations. These PC axes are then used as latent (hidden) factors in regression models, which test for associations between allele frequencies and the focal environmental variables - in our case latitude and longitude.

As a first step, we will test for clinality of the inversions along latitude and longitude in Europe and North America. We will fit general linear models including arcsine square-root transformed inversion frequencies as our dependent variables, which accounts for the skewed variance distribution in binomial data when normality is assumed. Then, we will overlay scatter-plots based on environmental variables and inversion frequencies with a logistic regression curve and print the *p*-value of the linear models in the top-right corner of each plot using the R-script Plot_Clinality.r.

```
## test for clinality of inversion frequency
for index in ${!DATA[@]}; do

    INVERSION=${DATA[index]}
    Rscript ${WD}/scripts/Plot_Clinality.r \
        ${INVERSION} \
        ${WD}
done
```

The plots in Figure 8 are consistent with our a-priori expectations. We do find highly significant inversion clines for both inversions along latitude in North America and, for *In(3R)Payne,* also in Europe. In all cases the inversions appear to be frequent in the South and to rapidly decline in frequency with increasing latitude. Conversely, very weak clines along longitude are only found for *In(3R)Payne* in Europe and for *In(2L)t* in North America. While these clines may be the result of spatially varying selection, we cannot rule out based on this simple analysis that the observed clinal patterns are alternatively the result of the demographic history that shaped the differentiation of the investigated populations.

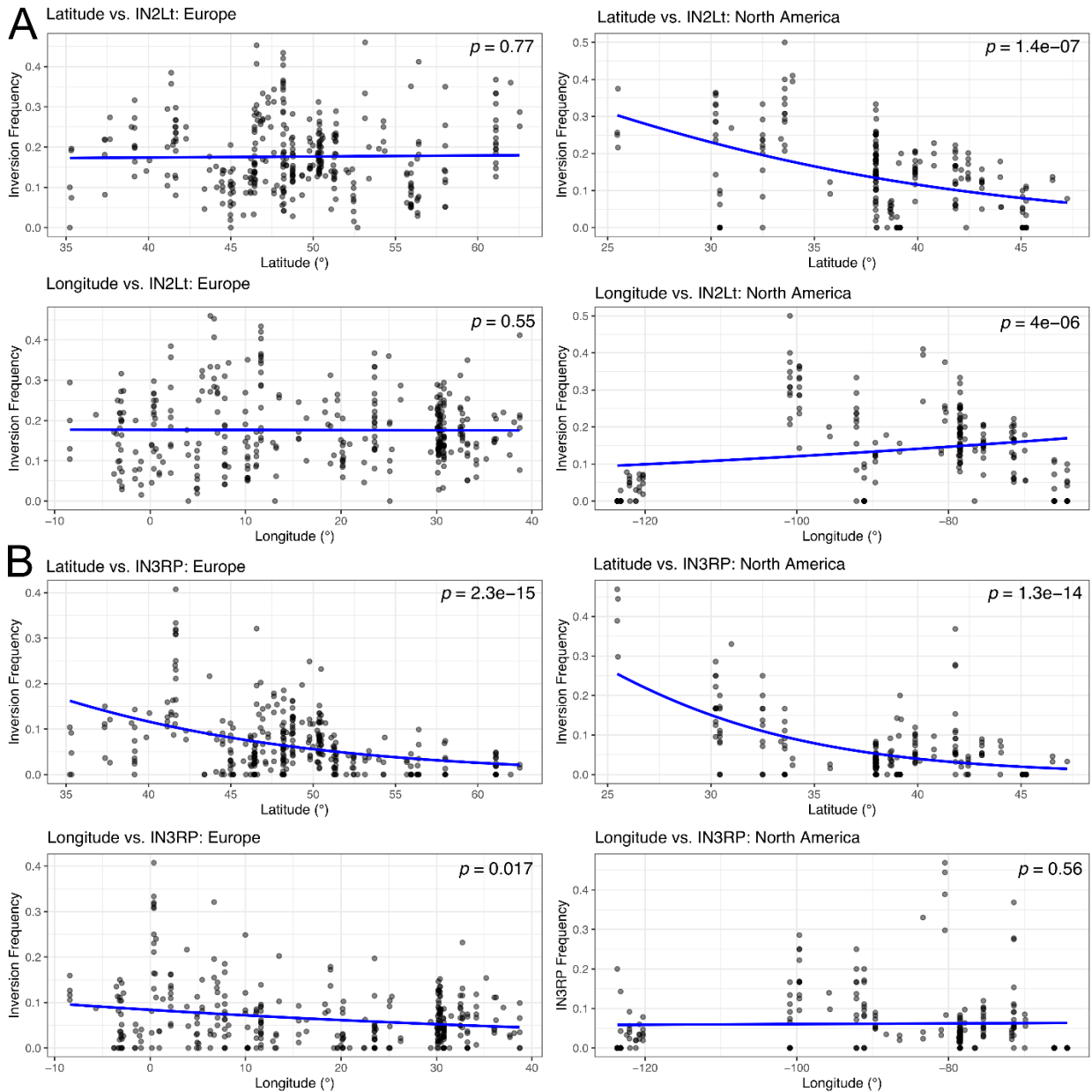

**Figure 8**: Scatterplots showing the association between inversion frequencies and both latitude (rows 1 and 3) and longitude (rows 2 and 4) in Europe (left) and North America (right) for *In(2L)t* (A) and *In(3R)Payne* (B). Logistic regression curves are shown in blue and the p-value of the regression models (adjusted $R^2$ value) are printed in the top-right corner of each plot.

In the final analyses, we will employ LFMMs to test if the inversion clines deviate from neutral expectations. Therefore, we will again use allele frequency matrices of the two continental subsets and add the frequency information of the inversion to the matrix. In addition, we will obtain information on latitudinal and longitudinal coordinates for all samples from the metadata table. Importantly, since we want to assess the clinal patterns of an inversion relative to the genomic background, we will exclude all SNPs located within the genomic region of the inversion for the calculation of the background to avoid confounding our analysis by a high degree of linkage with SNPs in this genomic region. Then, we will perform a PCA based on all SNPs and only consider a subset of the PC-axes as latent factors for the calculation of the latent factor mixed models. Here, we will include all PC-axes that cumulatively explain at least 25% of the genetic variation to avoid overfitting the model with too many latent factors. After calculating SNP-wise LFMMs using latitude and longitude as predictor variables, we will visualize the -$\log_{10}$-transformed $p$-values in Manhattan plots and include the $p$-value of the inversion as a horizontal bar at the corresponding genomic position. We will furthermore add the Bonferroni-corrected $p$-value threshold as a blue horizontal line. This threshold will be calculated by dividing the significance threshold $\alpha = 0.05$ by the total number of tests and will help us to account for a multiple testing problem which could lead to significant results by chance alone. Particularly, if the $p$-value of the inversion is higher than the -$\log_{10}$-transformed threshold, we can assume that the clinal patterns cannot be explained by neutral evolution alone. The R commands to carry out all the above-mentioned analyses steps are stored in LFMM.r.

```
### Test with LFMMs if clinality due to demography or potentially adaptive
for index in ${!DATA[@]}; do

    INVERSION=${DATA[index]}
    St=${Start[index]}
    En=${End[index]}
    Ch=${Chrom[index]}

    Rscript ${WD}/scripts/LFMM.r \
        ${INVERSION} \
        ${Ch} \
        ${St} \
        ${En} \
        ${WD}

done
```

The Manhattan plots in Figure 9 show that neither *In(2L)t* nor *In(3R)Payne* are significantly associated with latitude nor longitude in Europe, and are also not correlated with longitude in North America. However, both inversions exhibit highly significant correlations with latitude in North America, which indicates that these clinal patterns cannot be explained by the demographic history of the investigated populations alone. This finding is consistent with previous studies, and suggests that these two inversions presumably provide an advantage under certain environmental conditions at low latitudes but are not so beneficial in northern areas *[56, 61, 79, 86]*. Particularly the North American East coast, where most of the samples in our dataset were collected, is characterized by steep and continuous environmental gradients (such as temperature, precipitation and seasonality) ranging from subtropical conditions in southern Florida to temperate climates in Maine. Since many of these environmental gradients are highly intercorrelated, it is difficult to disentangle which factor influences clinal variation in inversion frequencies the most *[87]*. However, as a follow-up to the LFMM analysis based on latitude and longitude shown here, it would be worthwhile to focus on environmental variables, such temperature and precipitation, as predictors in similar downstream analyses to further explore which environmental conditions may influence the distribution of our two focal inversions the most.

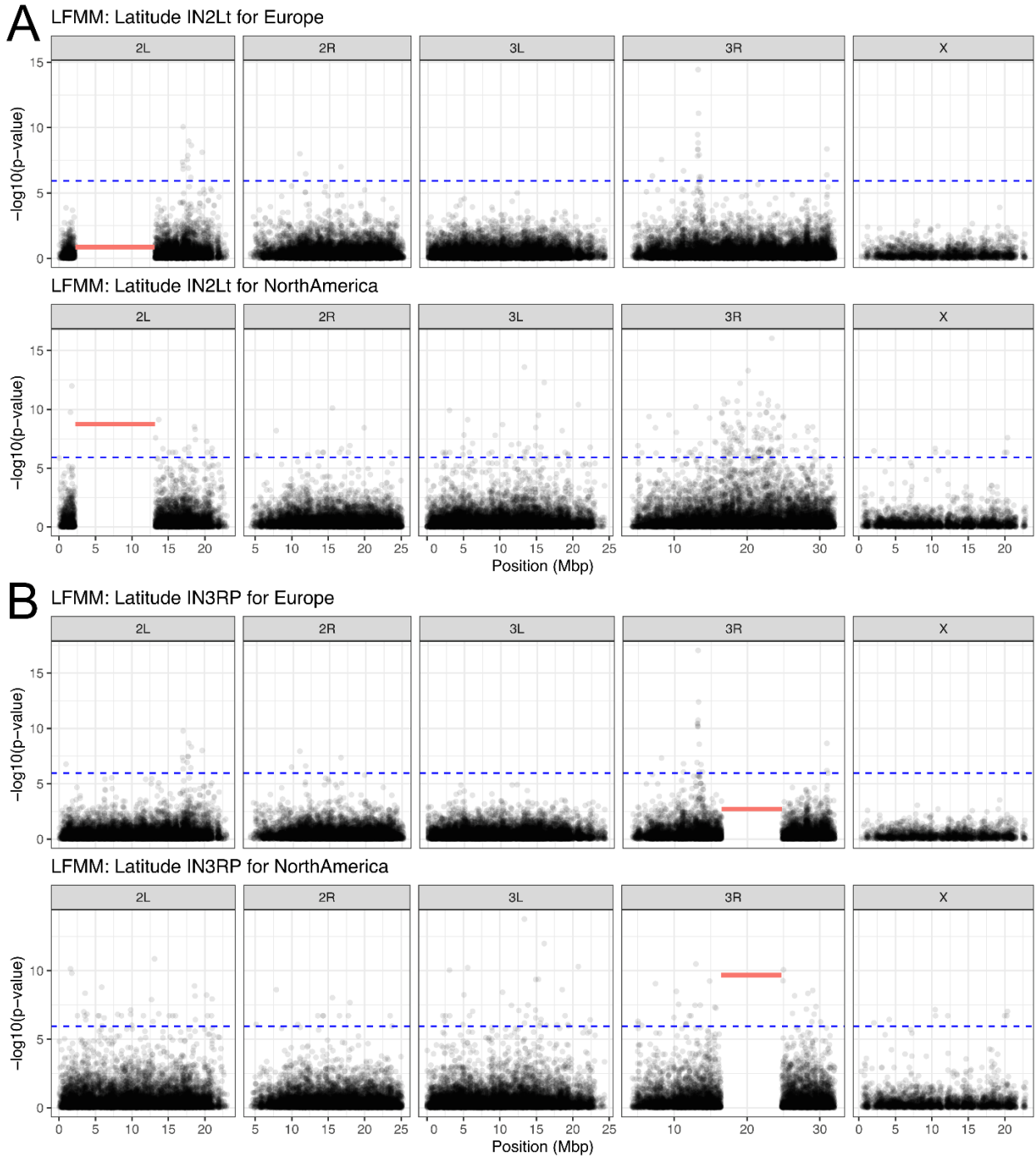

**Figure 9**: Manhattan plots showing the -log$_{10}$-transformed *p*-values of LFMM analyses that tested for significant associations between allele frequencies and environmental variables, in our case latitude and longitude (not shown). We included the inversion frequencies of *In(2L)t* (A) and *In(3R)Payne* (B) from Europe (rows 1 and 3) and North America (rows 2 and 4) in this analysis and show their *p*-value as a solid red line spanning their genomic region. In addition, we include the Bonferroni-corrected *p*-value threshold as a blue dashed line. Results for longitude are not shown since inversion frequencies were not significantly associated in any of the tests. However, the corresponding Manhattan plots can be found in the output folder in the GitHub repository.

## Acknowledgments

I am very grateful to Lino Ometto, who invited me to write this book chapter and to Elisabeth Haring for providing helpful comments to the manuscript. I further acknowledge the assistance of ChatGPT v. 4o, a language model developed by OpenAI, for helping with restructuring of the R and Python code that I developed for this project.

## Figure legends

**Figure 1**: Cytological preparation of polytene chromosomes from the salivary glands of a third instar *Drosophila melanogaster* larva which is heterozygote for the *In(3R)Payne* inversion on the right arm of the third chromosome (*3R*). The arrows pinpoint banding patterns that are diagnostic for the chromosomal arms, the dashed-lined ellipse indicates the inversion and the red and blue lines indicate the orientation of the two paired sister chromatids (inverted and standard, respectively) with different karyotype that form a characteristic inversion loop. The insert in the bottom right corner shows a schematic representation of the inversion loop, where the letters represents genes along the chromosome which should illustrate that the region spanned by "BCD" is inverted in the red haplotype as highlighted by a grey box on top of the linear alignment of the two karyotypes at the bottom of the insert.

**Figure 2**: Line plots showing the distribution of nucleotide diversity $\pi$ along the genome averaged in 200kbp non-overlapping windows for inverted (red) and non-inverted (blue individuals). The genomic regions spanned by the inversions *In(2L)t* (Panel A) and *In(3R)Payne* (Panel B) are highlighted by transparent blue boxes.

**Figure 3**: Manhattan plots showing the distribution of SNP-wise $F_{ST}$ between INV and ST individuals along the genome. In addition, averaged $F_{ST}$ values in 200kbp windows are overlaid as line plots in red and the genomic regions spanned by the inversions *In(2L)t* (A) and *In(3R)Payne* (B) are highlighted by transparent blue boxes.

**Figure 4**: Scatter plots showing the first two PC axes (Dim 1 and Dim 2) of PCAs based on SNPs in European (Panel A) or North American (Panel B) populations located either inside the genomic region spanned by an inversion (left two panels for *In(2L)t* and *In(3R)Payne*, respectively) or in the remaining genome (right-most panels). We highlight the country of origin using a combination of different colors and shapes. The numbers in parentheses next to the axis labels indicate the total variance explained by the corresponding PC-axis.

**Figure 5**: Frequencies of alleles from the diagnostic marker SNPs for *In(2Lt)* that are in strong LD plotted against their genomic position on chromosome 2L for one population sample collected in 2015 in Mautern, Lower Austria. A histogram of allele frequencies plotted on the right along the y-axis shows the range of frequencies across all marker SNPs. In addition, we plot the median allele frequency across all markers as a horizontal dashed line in red, which we consider the estimated inversion frequency in the population sample.

**Figure 6**: Scatterplots showing the association between inversion frequencies and the first PC-axis based on SNPs either located inside (left panels) our outside (right panels) the genomic region spanned by *In(2L)t* (A) or *In(3R)Payne* (B) for Europe (rows 2 and 4) and North America (rows 1 and 3). Regression lines based on linear regression are shown in blue and the determinants of correlation (adjusted $R^2$ value) are printed in the top-right corner of each plot.

**Figure 7**: Manhattan plots showing the results of SNP-wise logistic regressions that test for associations between inversion frequencies and allele frequencies at a given SNP. The *p*-values are -log10 - transformed and plotted along the y-axis. In addition, we highlight the genomic position of *In(2L)t* (A) and *In(3R)Payne* (B) by blue transparent boxes in Europe (rows 1 and 3) and North America (rows 2 and 4).

**Figure 8**: Scatterplots showing the association between inversion frequencies and both latitude (rows 1 and 3) and longitude (rows 2 and 4) in Europe (left) and North America (right) for *In(2L)t* (A) and *In(3R)Payne* (B). Logistic regression curves are shown in blue and the p-value of the regression models (adjusted R2 value) are printed in the top-right corner of each plot.

**Figure 9**: Manhattan plots showing the -$\log_{10}$-transformed *p*-values of LFMM analyses that tested for significant associations between allele frequencies and environmental variables, in our case latitude and longitude (not shown). We included the inversion frequencies of *In(2L)t* (A) and *In(3R)Payne* (B) from Europe (rows 1 and 3) and North America (rows 2 and 4) in this analysis and show their *p*-value as a solid red line spanning their genomic region. In addition, we include the Bonferroni-corrected *p*-value threshold as a blue dashed line. Results for longitude are not shown since inversion frequencies were not significantly associated in any of the tests. However, the corresponding Manhattan plots can be found in the output folder in the GitHub repository.